\documentclass[english,aps,amsmath,amssymb,superscriptaddress,twocolumn,reprint]{revtex4-1}
\usepackage{babel}
\usepackage{array}
\usepackage{multirow}
\usepackage{amsmath}
\usepackage{amssymb}
\usepackage{graphicx}
\def\br{{\bf r}}
\def\bp{{\bf r^\prime}}

\begin{document}

\title{Swift $GW$ beyond $10,000$ electrons using fractured stochastic
orbitals}

\author{Vojt\v{e}ch Vl\v{c}ek}
\email{vojtech.vlcek@gmail.com}

\affiliation{Department of Chemistry and Biochemistry, University of California,
Los Angeles California 90095, USA}

\affiliation{After July 1 2018: Department of Chemistry and Biochemistry, University
of California, Santa Barbara California 93106, USA}

\author{Wenfei Li}
\email{liwenfei@chem.ucla.edu}

\affiliation{Department of Chemistry and Biochemistry, University of California,
Los Angeles California 90095, USA}

\author{Roi Baer}
\email{roi.baer@huji.ac.il}

\affiliation{Fritz Haber Center for Molecular Dynamics, Institute of Chemistry,
The Hebrew University of Jerusalem, Jerusalem 91904, Israel}

\author{Eran Rabani}
\email{eran.rabani@berkeley.edu}

\affiliation{Department of Chemistry, University of California and Materials Science
Division, Lawrence Berkeley National Laboratory, Berkeley, California
94720, USA}

\affiliation{The Raymond and Beverly Sackler Center for Computational Molecular
and Materials Science, Tel Aviv University, Tel Aviv, Israel 69978}

\author{Daniel Neuhauser}
\email{dxn@ucla.edu}

\affiliation{Department of Chemistry and Biochemistry, University of California,
Los Angeles California 90095, USA}
\begin{abstract}
We introduce the concept of fractured stochastic orbitals (FSOs),
short vectors that sample a small number of space points and enable
an efficient stochastic sampling of any general function. As a first
demonstration, FSOs are applied in conjunction with simple direct-projection
to accelerate our recent stochastic $GW$ technique; the new developments
enable accurate prediction of $G_{0}W_{0}$ quasiparticle energies
and gaps for systems with up to $N_{e}>10,000$ electrons, with small
statistical errors of $\pm0.05\,{\rm eV}$ and using less than 2000
core CPU hours. Overall, stochastic $GW$ scales now linearly (and
often sub-linearly) with $N_{e}.$
\end{abstract}

\date{\today}
\maketitle

\section{Introduction}

Fundamental band gaps and quasiparticle (QP) energies determine the
electronic properties of molecules and solids, but their first principles
calculations are nontrivial. Density functional theory (DFT)~\cite{HohenbergKohn}
is usually used for ground state charge densities and atomic geometries,
but gives wrong QP energies.\cite{Dreizler1990,Martin2004,Aryasetiawan1998}
Going beyond DFT is computationally demanding. For small molecules,
configuration interaction \cite{shavitt1998history,sherrill1999configuration,martin2016interacting}
and the equation of motion coupled cluster technique \cite{rowe1968equations,stanton1993equation,krylov2008equation}
yield accurate QP energies, but scale very steeply with the number
of electrons.

In recent years, the $GW$ approximation~\cite{Hedin1965,Aryasetiawan1998,martin2016interacting}
became the predominant framework for QP calculations. The method describes
all many-body effects through the single-particle self-energy, approximated
as $\Sigma=GW$, where $G$ is the single particle Green's function
and $W$ is the screened Coulomb interaction. $GW$ provides accurate
ionization energies and electron affinities for both molecules and
solids, at a steep scaling.\cite{HybertsenLouie,umari2009optimal,umari2010gw,DeslippeSamsonidzeStrubbeEtAl2012,ljungberg2015cubic,govoni2015large,wilhelm2018toward}
Most computational improvements focus on reducing the prefactor rather
than lowering the overall scaling.\cite{umari2010gw,govoni2015large}

We recently introduced a stochastic formulation of $GW$ ~\cite{Neuhauser2014}
that expresses the self-energy as a statistical quantity, averaged
over random samplings. The resulting stochastic $GW$ method reproduces
the results of deterministic $GW$~\cite{vlcek2017stochastic} but
is very fast so it is applicable to large systems with thousands of
valance electrons.\cite{Neuhauser2014,Vlcek2016,vlcek2018quasiparticle} 

Here, two major improvements of stochastic $GW$ are introduced, and
together they enable routine calculations of QP energies for systems
with $N_{e}>10,000$. The first relates to the projection of random
functions to the occupied subspace. Originally, we used a Chebyshev
projection that is quite expensive. Here, we use a simpler direct
projection method that formally scales as $O(N_{e}^{2})$ but with
a small prefactor so it significantly reduces the overall effort (as
long as the occupied eigenstates or their linear combinations are
available). 

The second improvement relates to the conversion of causal potentials
to time-ordered potentials, which is a necessary ingredient in stochastic
$GW$. Originally, we used for this a stochastic basis (stochastic
resolution of the identity, S-RI), but it turns out that the required
number of stochastic basis functions grows with system size, destroying
the overall linear scaling for large systems. To circumvent this,
we develop a new approach based on short stochastic vectors, which
we label as fractured stochastic resolution of the identity (FS-RI);
the method does not lower the accuracy but is much cheaper, thereby
enabling the treatment of very large systems with $N_{e}>10,000$.
FS-RI has potentially a large number of applications, and we use stochastic
$GW$ here to demonstrate its efficiency. 

With direct projection and FS-RI, stochastic $GW$ is efficient and
scales very gently, as demonstrated here for finite molecules (acenes
and C$_{60}$ molecules) and periodic systems with large supercells. 

The paper is organized as follows: Deterministic $GW$ is reviewed
in Sec.$\ $II. In Sec.$\ $III we briefly explain (see Refs.~\onlinecite{Neuhauser2014} and \onlinecite{vlcek2017stochastic}
for details) how the stochastic expansion of $G$ converts $GW$ into
the action of $W$ on a source term. Sec.$\ $IV reviews the use of
linear response with deterministic or stochastic TDH (time-dependent
Hartree) for acting with $W^{R},$ the causal (retarded) effective
interaction. In Sec.$\ $V fractured orbitals are introduced and used
to convert the application of $W^{R}$ to $W$, and the overall algorithm
is reviewed in Sec.$\ $VI. Results for molecules and solids are shown
in Sec.$\ $VII, followed by conclusions in Sec.$\ $VIII.

\section{The GW method}

We first outline deterministic $GW$. The starting point is a specific
real-valued KS (Kohn-Sham) orbital $\phi$ (typically the HOMO or
LUMO) and associated eigenvalue $\varepsilon^{KS}$ that fulfill $H_{0}\phi=\varepsilon^{KS}\phi.$
Here the KS-DFT Hamiltonian is (using atomic units, and treating closed-shell
systems):

\[
H_{0}=-\frac{1}{2}\nabla^{2}+v_{{\rm nuc}}[n_{0}]+v_{{\rm H}}[n_{0}]+v_{{\rm xc}}[n_{0}],
\]
and we introduced the ground state density $(n_{0}(\br))$ and the
nuclear and exchange-correlation potentials, while the Hartree potential
is $v_{{\rm H}}[n](\br)=\int\nu\left(\br,\bp\right)n(\bp)d\bp$ with
$\nu\left(\br,\bp\right)=|\br-\bp|^{-1}.$ In the diagonal approximation,
the associated QP energy fulfills:\cite{Aryasetiawan1998}
\begin{align}
\varepsilon^{QP} & =\varepsilon^{KS}+\left\langle \phi\left|X+\Sigma\left(\omega=\varepsilon^{QP}\right)-v_{{\rm xc}}\right|\phi\right\rangle ,\label{eq:qpeq}
\end{align}
where $X$ is the Fock exchange-operator and $\Sigma$ refers throughout
to the polarization self-energy (rather than the full one). 

The frequency-resolved matrix element of the polarization self energy
is obtained from the time-dependent form, $\left\langle \phi\left|\Sigma(\omega)\right|\phi\right\rangle \equiv\int\left\langle \phi\left|\Sigma(t)\right|\phi\right\rangle e^{-\frac{1}{2}\gamma^{2}t^{2}}e^{i\omega t}dt$
where $\gamma$ is an energy broadening term for converging the time
integration.\footnote{Note that generally quantities in time and frequency use the same
symbol, and the specifics are clear from the argument.} The required polarization self-energy $\Sigma(t)$ has a very simple
direct product form in the $GW$ approximation:\cite{Aryasetiawan1998}

\begin{equation}
\Sigma(\br,\bp,t)=iG(\br,\bp,t)W(\br,\bp,t),\label{eq:Hedin}
\end{equation}
where $G$ is the Green's function (detailed below), and $W$ the
effective polarization interaction. We use here the one-shot $G_{0}W_{0}$
approximation, but omit the 0 subscript throughout, as well as the
$P$ (polarization) subscript on $\Sigma$ and $W$. Despite its elegance,
it is expensive to directly calculate $\left\langle \phi\left|\Sigma(t)\right|\phi\right\rangle $
using Eq. (\ref{eq:Hedin}) and the first goal of stochastic $GW$
is to convert the direct product to an initial value expression as
detailed below.

\section{Stochastic paradigm for resolving G }

\subsection{Resolution of identity}

Our starting point is a set of random functions on a grid, each labeled
$\bar{\zeta}(\br)$. The simplest choice is real discrete stochastic
functions that have a random sign at each point:

\[
\bar{\zeta}(\br)=\pm(dV)^{-\frac{1}{2}}
\]
($dV$ is the grid volume element). The stochastic functions fulfill
$\bar{\{\zeta}(\br)\bar{\zeta}(\bp)\}=(dV)^{-1}\delta_{\boldsymbol{\br,\bp}}$,
where $\delta_{\boldsymbol{\br,\bp}}$ is a Kronecker delta and the
$\left\{ \cdots\right\} $ refers to a statistical average over all
stochastic functions. This implies a resolution of the identity relation,
${\cal I}=\{|\bar{\zeta}\rangle\langle\bar{\zeta}|\}$. In practice
we need to use a finite number (labeled $N_{\bar{\zeta}})$ of stochastic
functions and the resolution becomes approximate
\begin{equation}
{\cal I}\simeq\frac{1}{N_{\bar{\zeta}}}\sum_{\bar{\zeta}}|\bar{\zeta}\rangle\langle\bar{\zeta}|.\label{eq:RI_approx}
\end{equation}

\subsection{Separable expression for the Green's function}

It is easy to show that the Kohn-Sham Green's function is given by
the operator form $iG(t)=e^{-iH_{0}t}\left(({\cal I}-{\cal P})\theta(t)-{\cal P}\theta(-t)\right)$
where the projection operator to the $N_{{\rm occ}}$ occupied states
is ${\cal P}=\sum_{n\le N_{{\rm occ}}}|\phi_{n}\rangle\langle\phi_{n}|$.
To make a separable form, multiply $iG(t)$ by Eq.~(\ref{eq:RI_approx}),
leading to: 

\begin{equation}
iG(\br,\bp,t)\simeq\frac{1}{N_{\bar{\zeta}}}\sum_{\bar{\zeta}}\zeta(\br,t)\bar{\zeta}(\bp),\label{eq:Gtexpansion}
\end{equation}
where $|\zeta(t)\rangle\equiv iG(t)|\bar{\zeta}\rangle$. Eq.~(\ref{eq:Gtexpansion})
is the main ingredient of stochastic $GW$, reformulating the Green's
function as a sum over separable terms. 

To evaluate $|\zeta(t)\rangle$, start with the negative-time Green's
function which is a propagator of the occupied states, $iG(t<0)=-e^{-iH_{0}t}{\cal P},$
so:

\begin{equation}
|\zeta(t<0)\rangle=-e^{-iH_{0}t}|\zeta^{v}\rangle,\label{eq:ztneg}
\end{equation}
where we define a stochastic occupied (valence) state $|\zeta^{v}\rangle={\cal P}|\bar{\zeta}\rangle.$
Similarly for positive times the Green's function is the propagator
of unoccupied (conduction) states, $iG(t>0)=e^{-iH_{0}t}({\cal I}-{\cal P}),$
so:

\begin{equation}
|\zeta(t>0)\rangle=e^{-iH_{0}t}|\zeta^{c}\rangle,\label{eq:ztpos}
\end{equation}
where $|\zeta^{c}\rangle=({\cal I}-{\cal P})|\bar{\zeta}\rangle=|\bar{\zeta}-\zeta^{v}\rangle.$

\subsection{Projective Filtering}

The next stage is therefore to calculate ${\cal P}|\bar{\zeta}\rangle$.
Previously we used Chebyshev filtering which scales linearly with
system size, but with a large prefactor. Therefore as long as the
occupied states are available (i.e., for systems with up to tens of
thousands of electrons) it is faster to use projective filtering,
i.e.,

\begin{equation}
\zeta^{v}(\br)=\langle\br|{\cal P}\bar{\zeta}\rangle=\sum_{n\le N_{{\rm occ}}}\phi_{n}(\br)\langle\phi_{n}|\bar{\zeta}\rangle.\label{eq:P-zetabar}
\end{equation}
In addition, the time-dependent orbitals of Eqs.~(\ref{eq:ztneg})
and (\ref{eq:ztpos}) are evaluated by a Trotter (split-operator)
propagation, $|\zeta(t\pm dt)\rangle=e^{\mp iH_{0}dt}|\zeta(t)\rangle,$
for positive or negative times respectively.

\subsection{Separable expression for $\langle\Sigma\rangle$}

Given Eq.~(\ref{eq:Hedin}) and the separable form of Eq.~(\ref{eq:Gtexpansion})
it immediately follows that 

\begin{equation}
\left\langle \phi\middle|\Sigma\left(t\right)\middle|\phi\right\rangle \simeq\frac{1}{N_{\bar{\zeta}}}\sum_{\bar{\zeta}}\int\phi\left(\br\right)\zeta\left(\br,t\right)u\left(\br,t\right)d\br,\label{eq:SigmaP_u}
\end{equation}
where

\begin{equation}
u\left(\br,t\right)=\int W\left(\br,\bp,t\right)\bar{\zeta}(\bp)\phi(\bp)d\bp.\label{eq:urt}
\end{equation}

\section{Acting with the retarded polarization potential}

To calculate $u\left(\br,t\right)$ in Eq.~(\ref{eq:urt}), one needs
to act with $W\left(\br,\bp,t\right)$ on the product $\bar{\zeta}(\bp)\phi(\bp)$.
This will be done in two stages: First, we will calculate the action
of the retarded (causal) effective-interaction:

\begin{equation}
u^{R}\left(\br,t\right)=\int W^{R}\left(\br,\bp,t\right)\bar{\zeta}(\bp)\phi(\bp)d\bp,\label{eq:uRt}
\end{equation}
and the next section explains how to convert the causal $u^{R}\left(\br,t\right)$
function to the time-ordered one $u\left(\br,t\right)$.

\subsection{\textmd{Deterministic $W^{R}$}}

It is well-known (Refs.~\onlinecite{martin2016interacting} and \onlinecite{hedin1999correlation})
that linear-response TDH can be used to calculate the action of $W^{R}$.
In our context, this amounts to perturbing all occupied states,

\[
\phi_{n}^{\lambda}(\br,t=0)=e^{-i\lambda v_{{\rm pert}}(\br)}\phi_{n}(\br),\,\,\,\,n\le N_{{\rm occ}}
\]
where $\lambda$ is small (typically $10^{-4}{\it E_{h}^{-{\rm 1}}})$
and $v_{{\rm pert}}(\br)\equiv\int\nu\left(\br,\bp\right)\bar{\zeta}(\bp)\phi(\bp)d\bp$.
Then one propagates simultaneously all occupied states, $|\phi_{n}^{\lambda}\left(t+dt\right)\rangle=e^{-iH^{\lambda}(t)dt}|\phi_{n}^{\lambda}\left(t\right)\rangle$
using a time-dependent Hamiltonian:

\begin{equation}
H^{\lambda}(t)=H_{0}+v_{{\rm H}}^{\lambda}(\br,t)-v_{{\rm H}}(\br),\label{eq:Hgammat}
\end{equation}
where $v_{{\rm H}}^{\lambda}(\br,t)\equiv v_{{\rm H}}[n^{\lambda}(t)](\br),$
$v_{{\rm H}}(\br)\equiv v_{{\rm H}}[n_{0}](\br)$ and 
\[
n^{\lambda}({\bf \br},t)=2\sum_{n\le N_{{\rm occ}}}|\phi_{n}^{\lambda}\left(\br,t\right)|^{2},
\]
where the density includes the spin factor. The causal response of
Eq.~(\ref{eq:uRt}) is then
\begin{equation}
u^{R}\left(\br,t\right)=\frac{v_{{\rm H}}^{\lambda}(\br,t)-v_{{\rm H}}(\br)}{\lambda}.\label{eq:urt_linresp}
\end{equation}

An alternative to this RPA screening procedure is to replace the TDH
by time-dependent DFT (TDDFT).\cite{runge1984density} In principle,
it is equivalent to the inclusion of a vertex function in $W$.\cite{del1994gwgamma,bruneval2005many}
Practical implementations with various approximate density functionals
revealed that this approach is not universally successful~\cite{del1994gwgamma,gruneis2014ionization}
but it often improves, at times dramatically, the energies of the
unoccupied states.\cite{Neuhauser2014,hung2016excitation}. Practically,
the only required changes are the replacement of all the Hartree potentials
in Eqs.~(\ref{eq:Hgammat}) and (\ref{eq:urt_linresp}) by the total
Hartree-exchange-correlation part, e.g., $v_{{\rm H}}^{\lambda}(\br,t)\to v_{{\rm H}}^{\lambda}(\br,t)+v_{{\rm xc}}[n^{\lambda}(t)](\br),$
etc. 

As a second alternative, the RPA form used here could be followed
by a zero-cost post processing self-consistency method, where a rigid
shift is applied on the Green's function part. This method improves
one-shot $G_{0}W_{0}$ and brings it to agreement with experiment;
see Ref.~\onlinecite{vlcek2017simple} for details.

\subsection{Stochastic $W^{R}$}

Deterministic linear-response TDH is expensive for large systems since
all occupied states are propagated. We have therefore developed and
applied a very cheap alternative, stochastic TDH.\cite{Neuhauser2014,Rabani2015,Rabani2015a}
For each $|\bar{\zeta}\rangle$ one chooses and propagates a small
set $(N_{\eta}\sim5-30)$ of occupied stochastic functions formally
defined as:

\begin{equation}
\eta_{l}(\br)=\sum_{n\le N_{{\rm occ}}}\eta_{nl}\phi_{n}(\br),\,\,\,\,l=1,...,N_{\eta},\label{eq:eta}
\end{equation}
where the coefficients can be real or complex, and are either specified
directly (e.g., $\eta_{nl}=\pm1)$ or based on a projection of a random
vector $\bar{\eta}_{l}(\br)$, i.e., $|\eta_{l}\rangle=P|\bar{\eta_{l}}\rangle$
(see Ref.~\onlinecite{neuhauser2015stochastic}). Then, completely
analogously to the deterministic case, the stochastic-occupied states
are perturbed 
\begin{equation}
\eta_{l}^{\lambda}(\br,t=0)=e^{-i\lambda v_{{\rm pert}}\left(\br\right)}\eta_{l}(\br),\label{eq:etaperturb}
\end{equation}
and propagated, 
\begin{equation}
|\eta_{l}^{\lambda}\left(t+dt\right)\rangle=e^{-iH^{\lambda}(t)dt}|\eta_{l}^{\lambda}\left(t\right)\rangle,\label{eq:etatpdt}
\end{equation}
and the time-dependent Hamiltonian is constructed again using Eq.
(\ref{eq:Hgammat}) but now the Hartree potential $v_{{\rm H}}^{\lambda}(\br,t)$
is based on the density of the propagated stochastic-occupied orbitals,

\begin{equation}
n^{\lambda}(\br,t)=C_{{\rm norm}}\frac{2}{N_{\eta}}\sum_{l\le N_{\eta}}\eta_{l}^{\lambda}({\bf \br},t)|^{2},
\end{equation}
where $C_{{\rm norm}}$ is a normalization constant ensuring the correct
total number of electrons ($\int$$n^{\lambda}(\br,t)d\br=N_{e}$). 

One last difference from the deterministic case is that it is necessary
now to repeat the calculation with $\lambda=0$ and the response is
then the difference of the time-dependent potentials

\begin{equation}
u^{R}\left(\br,t\right)=\frac{v_{{\rm H}}^{\lambda}(\br,t)-v_{{\rm H}}^{\lambda=0}(\br,t)}{\lambda}.\label{eq:uRt_stochW}
\end{equation}
Note that this is not needed in the deterministic case where $v_{{\rm H}}^{\lambda=0}(\br,t)=v_{{\rm H}}(\br)$
does not change in time; but even without perturbation the stochastic
TDDFT orbitals are not eigenstates and change in time leading to fluctuations
in the density, so Eq.~(\ref{eq:uRt_stochW}) is required to ensure
that the response is indeed in the linear regime.

\section{Fractured stochastic orbitals and the causal to time-ordered transformation}

$W$ and $W^{R}$ are related in Fourier space \textendash{} they
are the same for positive frequencies and are complex-conjugates at
negative frequencies.\cite{Fetter1971} The same properties are true
for $u$ and $u^{R}$, as long as the source term ($\bar{\zeta}\phi$)
in Eq. (\ref{eq:uRt}) is real. Practically, this gives a recipe for
obtaining $u$ from $u^{R},$ which we label as $u={\rm \mathcal{T}({\it u^{R}{\rm ),}}}$
meaning: FFT $u^{R}$ from time to frequency, conjugate at negative
frequencies and inverse FFT the result back to time 
\begin{align}
u^{R}(\br,t) & \to u^{R}(\br,\omega)=\int_{0}^{\infty}e^{-\frac{1}{2}\gamma^{2}t^{2}}e^{i\omega t}u^{R}(\br,t)dt\nonumber \\
 & \to u(\br,\omega)=\begin{cases}
u^{R}(\br,\omega) & \omega\ge0\\
\left(u^{R}(\br,\omega)\right){}^{*} & \omega<0
\end{cases}\label{eq:TO}\\
 & \to u(\br,t)=\frac{1}{2\pi}\int_{-\infty}^{\infty}e^{-i\omega t}u(\br,\omega)d\omega.\nonumber 
\end{align}
This procedure is, however, storage intensive since the whole $u^{R}(\br,t)$
from each core needs to be stored on disk.

\subsection{Stochastic basis}

Our previous approach (Ref.~\onlinecite{Neuhauser2014}) to solving
the storage issue was based on a stochastic resolution of identity,
Eq.~(\ref{eq:RI_approx}), 

\begin{equation}
u(\br,t)\simeq u_{{\rm aprx}}(\br,t)\equiv\frac{1}{N_{\xi}}\sum\xi(\br)u_{\xi}(t),\label{eq:uxi}
\end{equation}
where $\xi(\br)=\pm(dV)^{-0.5}$ is a third set of random functions
(beyond $\bar{\zeta}(\br)$ and $\eta(\br)$). Here $u_{\xi}(t)$
are obtained by time-ordering $\left(u_{\xi}={\rm \mathcal{T}{\it {\rm ({\it u_{\xi}^{R}})}}}\right)$
the causal coefficients $u_{\xi}^{R}(t)\equiv\lambda^{-1}\left(v_{\xi}^{\lambda}-v_{\xi}^{\lambda=0}\right)$
where $v_{\xi}^{\lambda}=\langle\xi|v_{{\rm H}}^{\lambda}(t)\rangle$
(see Eq. (\ref{eq:uRt_stochW})).

In the appendix, we prove that the relative error in the stochastic
expansion of $u$ (at a fixed time $t$) is the ratio of the number
of grid points and the number of stochastic vectors (cf., Eq. (\ref{eq:s2f_reltv})):
\[
\frac{\sigma^{2}\left(u(t)\right)}{\langle u(t)|u(t)\rangle}\equiv\frac{\left\{ \left\langle u_{{\rm aprx}}(t)-u(t)|u_{{\rm aprx}}(t)-u(t)\right\rangle \right\} }{\langle u(t)|u(t)\rangle}=\frac{N_{g}}{N_{\xi}}.
\]
This implies that the accuracy of the stochastic decreases with system
size, unless $N_{\xi}$ is increased. We previously (Refs.~\onlinecite{Neuhauser2014}
and \onlinecite{vlcek2017stochastic}) used $N_{\xi}=100-300$, but
for the very large systems studied here $N_{\xi}$ needs to be increased
to avoid large statistical errors. For large $N_{\xi}$, however,
the overlaps $\langle\xi|u^{R}(t)\rangle$ dominate the computational
cost, altering the linear scaling with system size. 

\subsection{Fractured basis:\label{subsec:Fractured-basis:}}

In order to overcome this drawback of the stochastic basis, we use
random functions in Eq.~(\ref{eq:uxi}) that are non-zero only over
short segments rather than extending over the full grid; we label
them as a ``fractured'' stochastic basis. 

A simple example clarifies this concept. Break the $N_{g}$ grid points
to two sets $A,\,B$, each with $N_{g}^{A}=N_{g}^{B}=\frac{1}{2}N_{g}$
points. Apply the stochastic resolution again with $N_{\xi}$ functions,
but now the first half of the functions ($\xi_{A})$ are non-zero
only over the $A$-set points, and the other half are non vanishing
over the $B$ set. Then, in an obvious notation:

\begin{equation}
u(\br,t)\simeq\begin{cases}
\frac{1}{N_{g}^{A}}\sum_{\xi_{A}}\xi_{A}(\br)u_{\xi,A}(t) & \br\in A\\
\frac{1}{N_{g}^{B}}\sum_{\xi_{B}}\xi_{B}(\br)u_{\xi,B}(t) & \br\in B,
\end{cases}\label{eq:u-example}
\end{equation}
where $u_{\xi,A}(t)\equiv\langle\xi_{A}|u\rangle_{A}\equiv dV\cdot\sum_{\br\in A}\xi_{A}(\br)u(\br)$
and analogously for $B$.

The cost of calculating each $u_{\xi,A}(t)$ is half that of calculating
the original $u_{\xi}(t)$, since the summation includes half the
grid points. But the squared standard deviation of $u$ is unchanged!

\begin{align*}
\sigma^{2}(u) & =\sigma_{A}^{2}(u)+\sigma_{B}^{2}(u)=\\
 & =\frac{N_{g}^{A}\langle u|u\rangle_{A}}{\frac{1}{2}N_{\xi}}+\frac{N_{g}^{B}\langle u|u\rangle_{B}}{\frac{1}{2}N_{\xi}}=\frac{N_{g}\langle u|u\rangle}{N_{\xi}},
\end{align*}
where we used $\langle u|u\rangle_{A}+\langle u|u\rangle_{B}=\langle u|u\rangle$.
This implies that the use of Eq.~\eqref{eq:uxi} instead of Eq.~\eqref{eq:u-example}
reduces the numerical effort by a factor of two without affecting
the statistical error.

Obviously, we could continue with this process of using smaller and
smaller segments further. In practice, we pick here a small segment
size $N_{{\rm {\it s}}}\sim0.001N_{g}-0.01N_{g}$, so that the ratio
of total grid length and the segment length, $L\equiv\frac{N_{g}}{N_{s}}$,
is $\approx1000$ . Each stochastic function now extends only over
$N_{s}$ points, so we label it as ``fractured''. For simplicity,
we do not even require the segments to be non-overlapping. The only
requirement is to ensure that each point has the same $L^{-1}$ probability
to be sampled, i.e., to have a fractured basis function that includes
it. \footnote{If a segment starting point is chosen near the first or last point
in the grid, then either the function should be wrapped (so a portion
of the segment is near the end of the grid and another portion is
near the beginning of the grid) or it should be padded (at the beginning
or end) with zeros, to guarantee that all points are equally sampled.} 

The fractured-stochastic basis expansion is then:

\begin{equation}
u(\br,t)\simeq\frac{L}{N_{\xi}}\sum_{{\rm }\xi\in{\rm frac}}\xi(\br)u_{\xi}(t),\label{eq:uFxi}
\end{equation}
where the ``frac'' label indicates that the summation extends over
fractured stochastic orbitals. Since each stochastic function $\xi(\br)$
is defined now only over $N_{s}$ points, the total cost in the expansion
is (for each time step) only $N_{s}N_{\xi}$, vs. $N_{g}N_{\xi}$
in the original stochastic expansion (Eq. (\ref{eq:uxi})). Therefore,
a much larger $N_{\xi}$ can now be used keeping the error in Eq.
(\ref{eq:uFxi}) in check.

We conclude this section by several observations:
\begin{enumerate}
\item The segments need to sufficiently sample each point; each grid point
has a probability $L^{-1}$ of being sampled by each of the $N_{\xi}$
functions so it is important to have $1\ll L^{-1}N_{\xi},$ i.e.,
$L\ll N_{\xi.}$. Put differently, the segment size cannot be too
small.
\item One could rewrite Eq.~(\ref{eq:uFxi}) as a formal fractured-stochastic
resolution of the identity, FS-RI:
\begin{equation}
{\cal I}\simeq\frac{L}{N_{\xi}}\sum_{\xi\in{\rm frac}}|\xi\rangle\langle\xi|.\label{eq:FS-RI}
\end{equation}
\item One could envision (although we have not done it here) that each segment
would be non-contiguous, i.e., made from $N_{s}$ random points from
the full grid. We do not even have to ensure that the points in each
segment are all different from each other, as long as they are randomly
selected!
\end{enumerate}

\section{Final Stochastic Algorithm}

The final stochastic GW algorithm is therefore simple:

Choose $N_{\bar{\zeta}}\sim200-1000$ stochastic functions (the wall
time is minimized if $N_{\bar{\zeta}}$ CPU cores are used, i.e.,
one per $\bar{\zeta}$). Then, for each choice of $\bar{\zeta}$: 
\begin{enumerate}
\item Choose a set of $N_{\xi}\sim5,000-50,000$ fractured random functions
$\xi(\br),$ each with $\frac{N_{g}}{L}$ grid points. Typically $L\sim100-1000$. 
\item Choose a set of $N_{\eta}\sim5-30$ stochastic-occupied functions
$\eta_{l}(\br)$ (Eq.~(\ref{eq:eta})). 
\item Calculate $v_{{\rm pert}}(\br)$ and perturb the $\eta_{l}(\br)$
per Eq.~(\ref{eq:etaperturb}). 
\item Propagate the perturbed $\eta_{l}^{\lambda}(\br,t)$ per Eq.~(\ref{eq:etatpdt}),
calculating at each time step $v_{{\rm H}}^{\lambda}(\br,t)$ and
constructing and storing in memory the set of $v_{\xi}^{\lambda}(t)$.
\item Repeat Step 4 for unperturbed functions (using $\lambda=0),$ storing
$v_{\xi}^{\lambda=0}(t)$ along the propagation. Then at the end of
the propagation calculate $u_{\xi}^{R}(t)$ and apply a time-ordering
operation $u_{\xi}={\rm \mathcal{T}({\it u_{\xi}^{R}})}$ (analogous
to Eq.~(\ref{eq:TO})).
\item Then calculate $\zeta(\br,t)$ for negative and positive times (Eqs.~(\ref{eq:ztneg})
and (\ref{eq:ztpos})) and use with Eq.~(\ref{eq:uFxi}) to accumulate
the matrix element of the self-energy (Eqs.~(\ref{eq:SigmaP_u})
and (\ref{eq:urt})). 
\end{enumerate}
Once steps 1-6 are finished average the resulting $\langle\phi|\Sigma(t)|\phi\rangle$
from each core, Fourier transform the result and solve Eq.~(\ref{eq:qpeq}). 

The algorithm above, using stochastic TDH, is the most efficient version
for large systems. If deterministic TDH is used, the steps are similar
except that instead of the stochastic occupied states $\eta_{l}$
one perturbs and propagates the deterministic occupied states $\phi_{n}(\br)$
(and then there is no need to calculate $v_{\xi}^{\lambda=0}(t)$,
which is obtained directly from the ground state density $n(\br)$).

\section{Simulations and results}

\label{sec:Implementation}

The stochastic $GW$ simulations were run on the Comet cluster with
Intel Xeon E5-2680v3 processors (2.5 GHz clock speed). The implementation
is trivially parallelized with speedup efficiency greater than $80\%$
when using up to $1728$ cores on $144$ CPUs. In all calculations
reported here all $12$ cores on each CPU were used. 

All simulations used uniform grids with isotropic spacing $dx=dy=dz$.
For both molecules and periodic solids, the KS-LDA ground state was
computed using Troullier-Martins pseudopotentials,\cite{TroullierMartins1991}
and a kinetic energy cutoff of 28~$E_{h}$. For molecules, the Martyna-Tuckerman
approach \cite{martyna1999reciprocal} was used to avoid the effect
of periodic images.

\subsection{Finite systems}

The new stochastic $GW$ implementation was first tested on acenes
with $1,2,4$ and $6$ rings as well as a ${\rm C}{}_{60}$ molecule.
Table~\ref{tab:finite_data} lists the parameters used for each system.
The uniform real-space grid spacing $dx$ is sufficiently small to
converge the LDA eigenvalues to $<10$~meV. Further, the QP shifts
are generally less sensitive to $dx$ than the LDA eigenvalues. The
damping parameter $\gamma$ cannot be too high to avoid over-broadening
the features in $\langle\phi|\Sigma\left(\omega\right)|\phi\rangle$.
For finite systems, $\gamma=0.1\,E_{{\rm h}}$ (cf., Eq.~\eqref{eq:TO})
was sufficient to converge $\varepsilon^{QP}$ (for a given $N_{\bar{\zeta}})$
to better than $0.01\,{\rm eV}$, although we used an even more conservative
value of $\gamma=0.06\,E_{{\rm h}}$.

\begin{figure}
\centering{} \includegraphics[width=1\columnwidth]{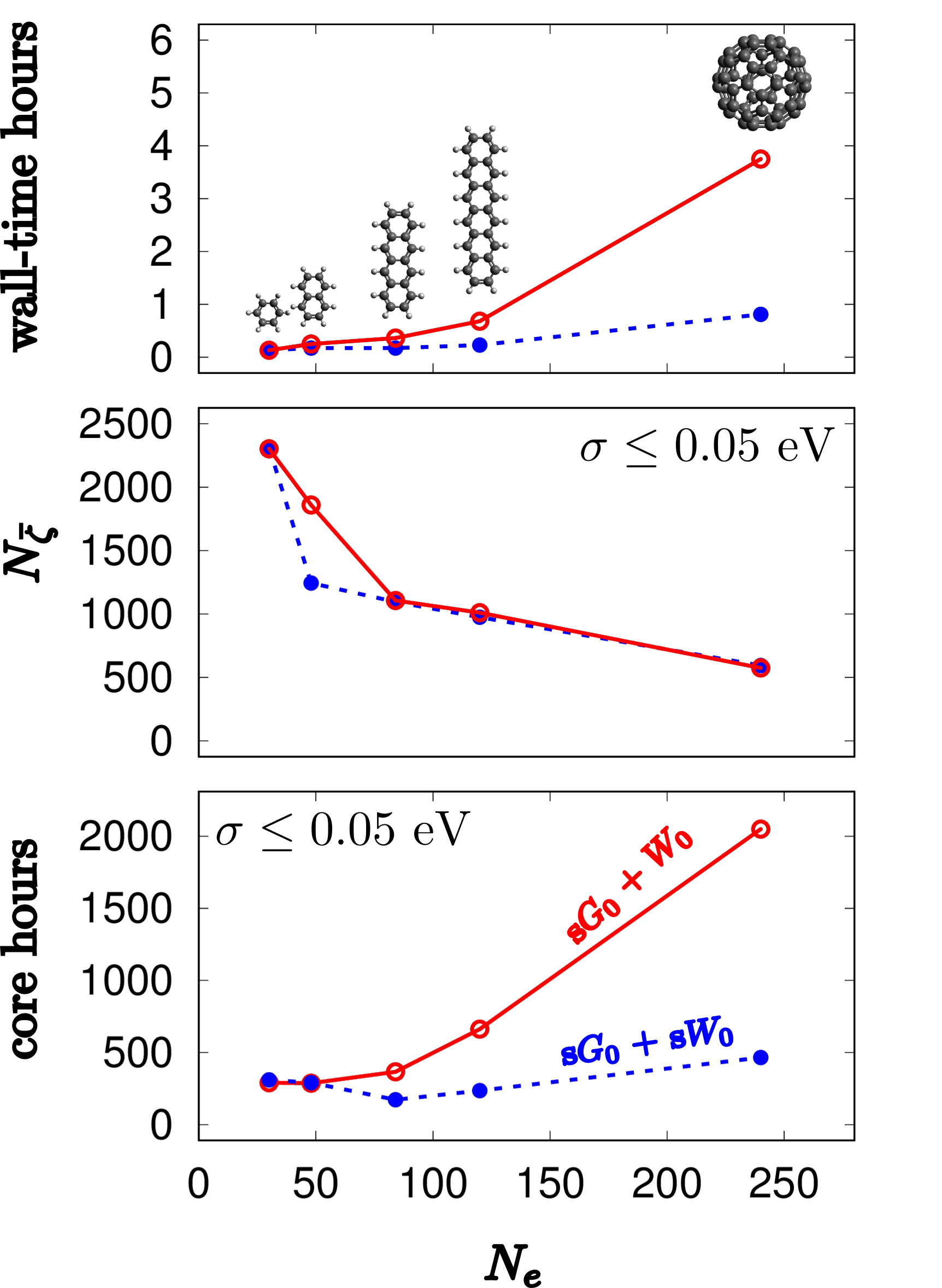}
\caption{\label{fig:finite-Nz-times}Resources needed to reduce the QP energy
stochastic error to $0.05{\rm \,eV}$ for acenes and ${\rm C_{60}}$.
Top panel: Wall time (hours); the dashed and solid lines refer to
deterministic and stochastic (with $N_{\eta}=16$) TDH propagations.
Middle panel: Number of stochastic vectors $N_{\bar{\zeta}}$; Bottom
panel: required CPU core hours. All calculations used $N_{\xi}=20,000$
and $L^{-1}=1\%$.}
\end{figure}

To isolate the influence of the number of stochastic TDH functions,
$N_{\eta}$, we studied the QP energies of the set of molecules with
deterministic and stochastic TDH propagation (the latter with $N_{\eta}=16).$
In both cases $N_{\bar{\zeta}}$ was increased till the resulting
statistical error for the HOMO and LUMO QP energies is $\le0.05$~eV.
Fig.~\ref{fig:finite-Nz-times} shows that the stochastic and deterministic
calculations require similar $N_{\bar{\zeta}}$, so the residual statistical
error due to the use of stochastic TDH is small.

The deterministic version scales quadratically with the size of the
system so as shown in Fig.~\ref{fig:finite-Nz-times} it quickly
becomes much more expensive than a constant-$N_{\eta}$ fully stochastic
treatment. Beyond tetracene the CPU time of the fully stochastic approach
(with a constant $N_{\eta}$) scales linearly with a slope of less
than 2 core-hours per electron.

Further, for large systems the number of propagated stochastic orbitals
$N_{\eta}$ can be reduced without increasing the stochastic error.
This is illustrated for C$_{60}$ where $N_{\eta}=8$ and $N_{\eta}=16$
(Table~\ref{tab:finite_data}) give an almost identical stochastic
error.

.

\begin{table}
\centering{}%
\begin{tabular}{|c|c|c|c|cc|cc|}
\hline 
System  & $N_{e}$  & $N_{g}$  & $N_{\eta}$  & \multicolumn{2}{c|}{HOMO } & \multicolumn{2}{c|}{LUMO}\tabularnewline
\hline 
\hline 
Benzene  & 30  & $(48)^{3}$ & 16  & -9.18  & $\pm$0.09  & 0.73  & $\pm$0.09\tabularnewline
\hline 
Naphtalene  & 48  & $48\cdot52\cdot60$ & 16  & -8.12  & $\pm$0.09  & -0.60  & $\pm$0.09 \tabularnewline
\hline 
Tetracene  & 84  & $48\cdot52\cdot82$ & 16  & -6.82  & $\pm$0.08  & -1.80  & $\pm$0.06 \tabularnewline
\hline 
Hexacene  & 120  & $48\cdot52\cdot104$ & 16  & -6.18  & $\pm$0.06  & -2.42  & $\pm$0.06 \tabularnewline
\hline 
\multirow{2}{*}{C$_{60}$ } & \multirow{2}{*}{240 } & \multirow{2}{*}{$(88)^{3}$} & 8  & -7.80  & $\pm$0.04  & -3.27  & $\pm$0.04 \tabularnewline
\cline{4-8} 
 &  &  & 16  & -7.78  & $\pm$0.04  & -3.30  & $\pm$0.04 \tabularnewline
\hline 
\end{tabular}\caption{\label{tab:finite_data}Estimated QP energies (eV) for a set of finite
systems with a fully stochastic approach. The calculations used $dx=0.35a_{0}$,
$N_{\bar{\zeta}}=750$, $N_{\xi}=20000$, and each fractured stochastic
function extended over only $L^{-1}=1\%$ of the total grid.}
\end{table}

\subsection{Periodic solids}

We next studied the performance of stochastic $GW$ for periodic systems
employing large real space grids (equivalent to $\Gamma$-point sampling
of large supercells in planewave codes). Specifically, we studied
the scaling of stochastic GW for silicon and diamond supercells including
several unit cells with lattice constants taken from experiment.\cite{yamanaka1996isotope,elliot2010structure}
The DFT eigenvalues were converged with respect to grid size to $<5$~meV,
with grid spacings of around $0.45$ and $0.35$ $a_{0}$ in all directions
for silicon and diamond, respectively. As in the molecular case, an
energy-broadening parameter of $\gamma=0.06\,{\rm {\it E_{{\rm h}}}}$
was sufficient for convergence.

Although the systems were large, most time was still spent on the
TDH stage. The initial projection and preparation of the stochastic
occupied orbitals, $\bar{\zeta}(\br)$ and $\eta_{l}(\br)$, took
at most $2\%$ of the CPU time. In addition, the FS-RI stage (converting
$u^{R}$ to $u$, Sec.$\,$\ref{subsec:Fractured-basis:}) took less
than $5\%$ of the total time when using $L=100$ (so each fractured
orbital covers only $1\%$ of the grid) and $N_{\xi}=20,000$. With
these parameters the component of the stochastic error in the QP shifts
due to the FS-RI is tiny, less than $0.01\,{\rm eV}$. 

We generally used $N_{\eta}=8$ propagated stochastic orbitals for
periodic systems. Higher values do not change the predicted QP energies
significantly, but reduce somewhat the statistical noise. When $N_{\eta}=16$
the fluctuations of $E_{g}$ in a 2$\times$2$\times$2 supercell
of diamond decrease by $8
$ (for the same $N_{\bar{\zeta}}$). This is not sufficient to offset
the cost (doubling the CPU time) of using $N_{\eta}=16$ so it is
it is better to fix $N_{\eta}=8$ and use a larger $N_{\bar{\zeta}}$.

\begin{figure}
\centering{}\includegraphics[width=1\columnwidth]{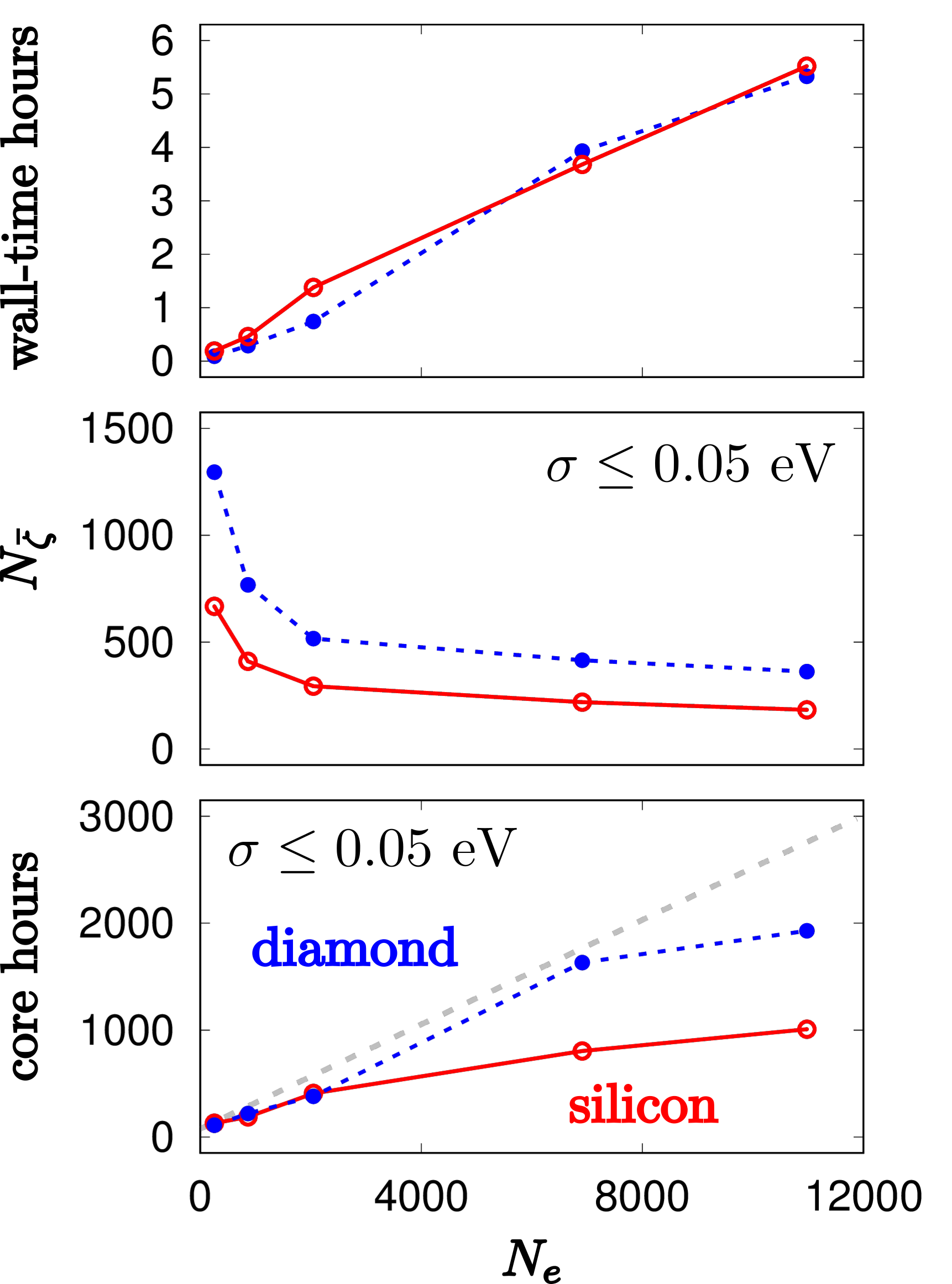}\caption{\label{fig:infinite-Nz-times}Resources required to lower the stochastic
error in $E_{g}$ to $0.05$~eV for silicon and diamond supercells
(red and blue, respectively). All calculations used $N_{\eta}=8$,
$N_{\xi}=20,000$ and $L^{-1}=1\%$. Top: CPU hours; Middle: Required
$N_{\bar{\zeta}}$ ; and bottom: Total CPU core hours.}
\end{figure}

Table~\ref{tab:solids_time}, obtained with a fixed $N_{\bar{\zeta}}=400$,
shows that the stochastic error of $E_{g}$ (the gap between the bottom
of the conduction band and the top of the valance band) decreases
rapidly with system size. Further, the number of stochastic vectors
$N_{\bar{\zeta}}$ required to decrease the error below 0.05~eV is
plotted in Fig.~\ref{fig:infinite-Nz-times}. The lower panel shows
that the total CPU time then scales at worst linearly with $N_{e}$.
The initial slope (fitted to the four smallest systems) is 0.25 core
hours per electron. The time to solution then quickly declines for
larger supercells as the required $N_{\bar{\zeta}}$ decreases. For
the largest supercells of both systems, we observe a linear slope
of $0.06$ core hours per electron. Specifically, calculations for
diamond and silicon supercells with $10978$ valence electrons consumed
only about $1900$ and $1000$ core hours!

Per-electron the periodic calculations were much faster (up to almost
20 times!) than for finite systems. One obvious reason is that it
is much easier to pack electrons in a periodic system, so, for example,
the largest supercell of silicon or diamond has 50 times more electrons
than ${\rm C_{60}}$ but its grid is only $\sim4$ times bigger. In
addition, the large periodic systems have many more electrons so they
required fewer samples $(N_{\bar{\zeta}})$.

\begin{table}
\centering{}%
\begin{tabular}{|c|c|c|c|c|c|c|}
\hline 
\multirow{2}{*}{$N_{cells}$} & \multirow{2}{*}{$N_{e}$} & \multirow{2}{*}{$N_{g}$} & \multicolumn{4}{c|}{$E_{g}\left(eV\right)$}\tabularnewline
\cline{4-7} 
 &  &  & \multicolumn{2}{c|}{Diamond} & \multicolumn{2}{c|}{Silicon}\tabularnewline
\hline 
\hline 
8  & 256  & $(42)^{3}$ & 5.36  & $\pm$0.09  & 1.17  & $\pm$0.06 \tabularnewline
\hline 
27  & 864  & $(60)^{3}$ & 5.28  & $\pm$0.07  & 1.35  & $\pm$0.05 \tabularnewline
\hline 
64  & 2048  & $(80)^{3}$ & 5.40  & $\pm$0.06  & 1.29  & $\pm$0.04 \tabularnewline
\hline 
216  & 6912  & $(120)^{3}$ & 5.55  & $\pm$0.04  & 1.24  & $\pm$0.04 \tabularnewline
\hline 
343  & 10978  & $(140)^{3}$ & 5.51  & $\pm$0.04  & 1.24  & $\pm$0.03 \tabularnewline
\hline 
\end{tabular}\caption{\label{tab:solids_time}Estimated QP gaps for bulk carbon and silicon
using $N_{\bar{\zeta}}=400$, $N_{\eta}=8$, $N_{\xi}=20,000$ and
$L=100$. $N_{{\rm {\it cells}}}$ is the number of conventional cells
in a supercell, $N_{e}$ the total number of valence electrons, and
$N_{g}$ is the total number of grid points. }
 
\end{table}

\section{Discussion and conclusions}

In conclusion, we introduced a general method for efficient stochastic
sampling, fragmented stochastic resolution of the identity, (FS-RI).
Here, we applied FS-RI to enhance our stochastic-$GW$ method. When
combined with a simple direct projection approach to efficiently obtain
random occupied orbitals from initial white noise vectors, the overall
stochastic $GW$ method is very fast, scales practically linearly,
and makes it possible to calculate QP energies for systems with $N_{e}>10,000$
valence electrons in only a few thousands of CPU-core hours or less.

The overall algorithm is straightforward, and an open-source software
(StochasticGW) is freely available.\footnote{The StochasticGW code is available under GPL at http://www.stochasticgw.com}
Our calculations show very favorable scaling of the statistical error
in all three types of stochastic samplings used in stochastic $GW$: 
\begin{itemize}
\item FS-RI makes it possible to easily increase the number of number of
such sampling vectors ($N_{\xi})$ by 100-fold or more, from hundreds
to tens of thousands. The key feature is that the accuracy is independent
of the size of the fractured stochastic vectors as long as each grid
point is sufficiently sampled (i.e., as long as $L\ll N_{\xi}$).
The FS-RI expansion (Eq. (\ref{eq:uFxi})) adds only a tiny stochastic
error (less than $0.01{\rm \,eV}$) and its cost is negligible. 
\item Very few propagated stochastic orbitals $\eta$ are needed for the
TDH propagation \textendash{} we used $N_{\eta}=16$ for molecules
but even half that number, $N_{\eta}=8$, was sufficient for large
molecules and periodic solids. 
\item The stochastic error depends on the number of vectors used to sample
the Green's function, $N_{\bar{\zeta}}$. To obtain a low error of
$0.05{\rm \,eV}$ in the quasiparticle energies, $N_{\bar{\zeta}}$
is circa 1000 for small systems but decreases with system size so
for ${\rm C}_{60}$ it is only 600 and for large periodic supercells
it decreases to a few hundreds. Our calculations here and in Ref.$\ $\cite{vlcek2018quasiparticle}
indicate that the stochastic fluctuations somewhat increases with
$E_{g}$, but linear scaling is maintained. 
\end{itemize}
Taken together, we find a very favorable scaling. Cells with 10978
valence electrons require less than 2000 core hours to yield QP energies with statistical errors below $0.05{\rm \,eV}$.
Our method thus makes it possible to calculate QP energies of extremely
large systems with thousands of atoms on small computer clusters. 

While our stochastic GW has a practically linear scaling wall-time,
it has two ingredients which formally scale non-linearly. We use occupied-projection,
which scales as ${\rm O({\it N_{e}^{2}{\rm )};}}$ this by itself
however is not a major issue since it will not be the dominant part
of the calculation until we would reach $N_{e}\gg100,000$. But more
importantly, occupied-projection uses the occupied DFT eigenstates,
and in most DFT codes the extraction of these states scales as ${\rm O({\it N_{e}^{{\it 3}}})}$
and is prohibitive for very large systems. We therefore anticipate
that when simulating systems with $N_{e}>50,000$ it may be necessary
to switch back to Chebyshev-projection that avoids the eigenstates
altogether, as long as the underlying DFT potential could be obtained
by either linear scaling DFT \cite{yang1991direct,hernandez1995self,mohr2015accurate,vandevondele2012linear}
or stochastic DFT.\cite{BaerNeuhauserRabani2013,neuhauser2014communication}

Finally, we note that the new technique invented in this paper, FS-RI,
is potentially useful for a large number of applications that are
unrelated to stochastic $GW$, including long-range exchange, stochastic
MP2 (direct and exchange), and stochastic resolution of the identity.\cite{neuhauser2015stochastic,neuhauser2017stochastic,takeshita2017stochastic} 
\begin{acknowledgments}
We are grateful for support by the Center for Computational Study
of Excited State Phenomena in Energy Materials (C2SEPEM) at the Lawrence
Berkeley National Laboratory, which is funded by the U.S. Department
of Energy, Office of Science, Basic energy Sciences, Materials Sciences
and Engineering Division under contract No. DEAC02-05CH11231 as part
of the Computational materials Sciences Program. V. V. greatly appreciates
helpful discussion with Gabriel Kotliar and Mark Hybertsen. The calculations
were performed as part of the XSEDE computational Project No. TG-CHE170058.\cite{towns2014xsede} 
\end{acknowledgments}

\appendix*
\setcounter{equation}{0}

\section*{Appendix: Statistical error of a stochastic basis expansion}

Given a stochastic expansion of a general function, analogous to Eq.~(\ref{eq:uxi}), 

\begin{equation}
|f\rangle\simeq|f_{{\rm aprx}}\rangle\equiv\frac{1}{N_{\xi}}\sum|\xi\rangle\langle\xi|f\rangle,\label{eq:fxi}
\end{equation}
we show here that the average relative error in the representation
of $f$ is proportional to the number of grid points. Specifically,
define
\begin{align}
\sigma^{2}\left(f\right) & =\left\{ \langle f_{{\rm aprx}}|f_{{\rm aprx}}\rangle\right\} -\langle f|f\rangle\nonumber \\
 & =\frac{1}{N_{\xi}^{2}}\left\{ \sum_{\xi,\xi'}\langle\xi|\xi'\rangle\langle f|\xi\rangle\langle\xi'|f\rangle\right\} -\langle f|f\rangle,
\end{align}
where all functions are assumed real. Separating yields
\begin{equation}
\sigma^{2}\left(f\right)=J_{1}+J_{2}-\langle f|f\rangle.
\end{equation}
Here $J_{1}$ is the $\xi=\xi'$ contribution

\begin{align*}
J_{1} & =\frac{1}{N_{\xi}^{2}}\left\{ \sum_{\xi'=\xi}\langle\xi|\xi\rangle\langle f|\xi\rangle\langle\xi|f\rangle\right\} \\
 & =\frac{N_{g}}{N_{\xi}^{2}}\left\{ \sum_{\xi}\langle f|\xi\rangle\langle\xi|f\rangle\right\} =\frac{N_{g}}{N_{\xi}^{2}}\left\{ \sum_{\xi}\langle f|\xi\rangle\langle\xi|f\rangle\right\} ,
\end{align*}
where the definition $\xi(\br)=\pm(dV)^{-0.5}$ implies that $\langle\xi|\xi\rangle=N_{g}$
(always, not just as an average). The resulting expression for $J_{1}$
simply involves a resolution of the identity $\left\{ |\xi\rangle\langle\xi|\right\} =I,$
so 

\begin{equation}
J_{1}=\frac{N_{g}}{N_{\xi}^{2}}\sum_{\xi}\langle f|f\rangle=\frac{N_{g}}{N_{\xi}^{2}}N_{\xi}\langle f|f\rangle.
\end{equation}
Similarly, $J_{2}$ is the $\xi'\ne\xi$ contribution

\[
J_{2}=\frac{1}{N_{\xi}^{2}}\left\{ \sum_{\xi}\sum_{\xi'\ne\xi}\langle\xi|\xi'\rangle\langle\xi'|f\rangle\langle f|\xi\rangle\right\} ,
\]
and since the condition $\xi'\ne\xi$ does not restrict $\xi',$ the
resolution of the identity $I=\left\{ |\xi'\rangle\langle\xi'|\right\} $
is still valid, so

\begin{align*}
J_{2} & =\frac{1}{N_{\xi}^{2}}\left\{ \sum_{\xi}\sum_{\xi'\ne\xi}\langle\xi|f\rangle\langle f|\xi\rangle\right\} \\
 & =\frac{1}{N_{\xi}^{2}}\sum_{\xi}\sum_{\xi'\ne\xi}\langle f|f\rangle=\frac{\langle f|f\rangle}{N_{\xi}^{2}}N_{\xi}\left(N_{\xi}-1\right).
\end{align*}
Adding the terms gives

\begin{equation}
\frac{\sigma^{2}(f)}{\langle f|f\rangle}=\frac{(N_{g}-1)}{N_{\xi}}\simeq\frac{N_{g}}{N_{\xi}},\label{eq:s2f_reltv}
\end{equation}
as stipulated.

\bibliographystyle{apsrev4-1}
\bibliography{fastGW}

\begin{thebibliography}{49}%
\makeatletter
\providecommand \@ifxundefined [1]{%
 \@ifx{#1\undefined}
}%
\providecommand \@ifnum [1]{%
 \ifnum #1\expandafter \@firstoftwo
 \else \expandafter \@secondoftwo
 \fi
}%
\providecommand \@ifx [1]{%
 \ifx #1\expandafter \@firstoftwo
 \else \expandafter \@secondoftwo
 \fi
}%
\providecommand \natexlab [1]{#1}%
\providecommand \enquote  [1]{``#1''}%
\providecommand \bibnamefont  [1]{#1}%
\providecommand \bibfnamefont [1]{#1}%
\providecommand \citenamefont [1]{#1}%
\providecommand \href@noop [0]{\@secondoftwo}%
\providecommand \href [0]{\begingroup \@sanitize@url \@href}%
\providecommand \@href[1]{\@@startlink{#1}\@@href}%
\providecommand \@@href[1]{\endgroup#1\@@endlink}%
\providecommand \@sanitize@url [0]{\catcode `\\12\catcode `\$12\catcode
  `\&12\catcode `\#12\catcode `\^12\catcode `\_12\catcode `\%12\relax}%
\providecommand \@@startlink[1]{}%
\providecommand \@@endlink[0]{}%
\providecommand \url  [0]{\begingroup\@sanitize@url \@url }%
\providecommand \@url [1]{\endgroup\@href {#1}{\urlprefix }}%
\providecommand \urlprefix  [0]{URL }%
\providecommand \Eprint [0]{\href }%
\providecommand \doibase [0]{http://dx.doi.org/}%
\providecommand \selectlanguage [0]{\@gobble}%
\providecommand \bibinfo  [0]{\@secondoftwo}%
\providecommand \bibfield  [0]{\@secondoftwo}%
\providecommand \translation [1]{[#1]}%
\providecommand \BibitemOpen [0]{}%
\providecommand \bibitemStop [0]{}%
\providecommand \bibitemNoStop [0]{.\EOS\space}%
\providecommand \EOS [0]{\spacefactor3000\relax}%
\providecommand \BibitemShut  [1]{\csname bibitem#1\endcsname}%
\let\auto@bib@innerbib\@empty
\bibitem [{\citenamefont {Hohenberg}\ and\ \citenamefont
  {Kohn}(1964)}]{HohenbergKohn}%
  \BibitemOpen
  \bibfield  {author} {\bibinfo {author} {\bibfnamefont {P.}~\bibnamefont
  {Hohenberg}}\ and\ \bibinfo {author} {\bibfnamefont {W.}~\bibnamefont
  {Kohn}},\ }\href@noop {} {\bibfield  {journal} {\bibinfo  {journal} {Phys.
  Rev.}\ }\textbf {\bibinfo {volume} {136}},\ \bibinfo {pages} {864} (\bibinfo
  {year} {1964})}\BibitemShut {NoStop}%
\bibitem [{\citenamefont {Dreizler}\ and\ \citenamefont
  {Gross}(1990)}]{Dreizler1990}%
  \BibitemOpen
  \bibfield  {author} {\bibinfo {author} {\bibfnamefont {R.~M.}\ \bibnamefont
  {Dreizler}}\ and\ \bibinfo {author} {\bibfnamefont {E.~K.~U.}\ \bibnamefont
  {Gross}},\ }\href@noop {} {\emph {\bibinfo {title} {Density Functional
  Theory: An Approach to the Quantum Many-Body Problem}}}\ (\bibinfo
  {publisher} {Springer Science \& Business Media},\ \bibinfo {year}
  {1990})\BibitemShut {NoStop}%
\bibitem [{\citenamefont {Martin}(2004)}]{Martin2004}%
  \BibitemOpen
  \bibfield  {author} {\bibinfo {author} {\bibfnamefont {R.~M.}\ \bibnamefont
  {Martin}},\ }\href {https://books.google.com/books?id=dmRTFLpSGNsC&pgis=1}
  {\emph {\bibinfo {title} {{Electronic Structure: Basic Theory and Practical
  Methods}}}}\ (\bibinfo  {publisher} {Cambridge University Press},\ \bibinfo
  {year} {2004})\ p.\ \bibinfo {pages} {624}\BibitemShut {NoStop}%
\bibitem [{\citenamefont {Aryasetiawan}\ and\ \citenamefont
  {Gunnarsson}(1998)}]{Aryasetiawan1998}%
  \BibitemOpen
  \bibfield  {author} {\bibinfo {author} {\bibfnamefont {F.}~\bibnamefont
  {Aryasetiawan}}\ and\ \bibinfo {author} {\bibfnamefont {O.}~\bibnamefont
  {Gunnarsson}},\ }\href {\doibase 10.1088/0034-4885/61/3/002} {\bibfield
  {journal} {\bibinfo  {journal} {Reports Prog. Phys.}\ }\textbf {\bibinfo
  {volume} {61}},\ \bibinfo {pages} {237} (\bibinfo {year} {1998})}\BibitemShut
  {NoStop}%
\bibitem [{\citenamefont {Shavitt}(1998)}]{shavitt1998history}%
  \BibitemOpen
  \bibfield  {author} {\bibinfo {author} {\bibfnamefont {I.}~\bibnamefont
  {Shavitt}},\ }\href@noop {} {\bibfield  {journal} {\bibinfo  {journal} {Mol.
  Phys.}\ }\textbf {\bibinfo {volume} {94}},\ \bibinfo {pages} {3} (\bibinfo
  {year} {1998})}\BibitemShut {NoStop}%
\bibitem [{\citenamefont {Sherrill}\ and\ \citenamefont
  {Schaefer~III}(1999)}]{sherrill1999configuration}%
  \BibitemOpen
  \bibfield  {author} {\bibinfo {author} {\bibfnamefont {C.~D.}\ \bibnamefont
  {Sherrill}}\ and\ \bibinfo {author} {\bibfnamefont {H.~F.}\ \bibnamefont
  {Schaefer~III}},\ }in\ \href@noop {} {\emph {\bibinfo {booktitle} {Advances
  in quantum chemistry}}},\ Vol.~\bibinfo {volume} {34}\ (\bibinfo  {publisher}
  {Elsevier},\ \bibinfo {year} {1999})\ pp.\ \bibinfo {pages}
  {143--269}\BibitemShut {NoStop}%
\bibitem [{\citenamefont {Martin}\ \emph {et~al.}(2016)\citenamefont {Martin},
  \citenamefont {Reining},\ and\ \citenamefont
  {Ceperley}}]{martin2016interacting}%
  \BibitemOpen
  \bibfield  {author} {\bibinfo {author} {\bibfnamefont {R.~M.}\ \bibnamefont
  {Martin}}, \bibinfo {author} {\bibfnamefont {L.}~\bibnamefont {Reining}}, \
  and\ \bibinfo {author} {\bibfnamefont {D.~M.}\ \bibnamefont {Ceperley}},\
  }\href@noop {} {\emph {\bibinfo {title} {Interacting Electrons}}}\ (\bibinfo
  {publisher} {Cambridge University Press},\ \bibinfo {year}
  {2016})\BibitemShut {NoStop}%
\bibitem [{\citenamefont {Rowe}(1968)}]{rowe1968equations}%
  \BibitemOpen
  \bibfield  {author} {\bibinfo {author} {\bibfnamefont {D.}~\bibnamefont
  {Rowe}},\ }\href@noop {} {\bibfield  {journal} {\bibinfo  {journal} {Rev.
  Mod. Phys.}\ }\textbf {\bibinfo {volume} {40}},\ \bibinfo {pages} {153}
  (\bibinfo {year} {1968})}\BibitemShut {NoStop}%
\bibitem [{\citenamefont {Stanton}\ and\ \citenamefont
  {Bartlett}(1993)}]{stanton1993equation}%
  \BibitemOpen
  \bibfield  {author} {\bibinfo {author} {\bibfnamefont {J.~F.}\ \bibnamefont
  {Stanton}}\ and\ \bibinfo {author} {\bibfnamefont {R.~J.}\ \bibnamefont
  {Bartlett}},\ }\href@noop {} {\bibfield  {journal} {\bibinfo  {journal} {J.
  Chem. Phys.}\ }\textbf {\bibinfo {volume} {98}},\ \bibinfo {pages} {7029}
  (\bibinfo {year} {1993})}\BibitemShut {NoStop}%
\bibitem [{\citenamefont {Krylov}(2008)}]{krylov2008equation}%
  \BibitemOpen
  \bibfield  {author} {\bibinfo {author} {\bibfnamefont {A.~I.}\ \bibnamefont
  {Krylov}},\ }\href@noop {} {\bibfield  {journal} {\bibinfo  {journal} {Annu.
  Rev. Phys. Chem.}\ }\textbf {\bibinfo {volume} {59}} (\bibinfo {year}
  {2008})}\BibitemShut {NoStop}%
\bibitem [{\citenamefont {Hedin}(1965)}]{Hedin1965}%
  \BibitemOpen
  \bibfield  {author} {\bibinfo {author} {\bibfnamefont {L.}~\bibnamefont
  {Hedin}},\ }\href {\doibase 10.1103/PhysRev.139.A796} {\bibfield  {journal}
  {\bibinfo  {journal} {Phys. Rev.}\ }\textbf {\bibinfo {volume} {139}},\
  \bibinfo {pages} {A796} (\bibinfo {year} {1965})}\BibitemShut {NoStop}%
\bibitem [{\citenamefont {Hybertsen}\ and\ \citenamefont
  {Louie}(1986)}]{HybertsenLouie}%
  \BibitemOpen
  \bibfield  {author} {\bibinfo {author} {\bibfnamefont {M.~S.}\ \bibnamefont
  {Hybertsen}}\ and\ \bibinfo {author} {\bibfnamefont {S.~G.}\ \bibnamefont
  {Louie}},\ }\href@noop {} {\bibfield  {journal} {\bibinfo  {journal} {Phys.
  Rev. B}\ }\textbf {\bibinfo {volume} {34}},\ \bibinfo {pages} {5390}
  (\bibinfo {year} {1986})}\BibitemShut {NoStop}%
\bibitem [{\citenamefont {Umari}\ \emph {et~al.}(2009)\citenamefont {Umari},
  \citenamefont {Stenuit},\ and\ \citenamefont {Baroni}}]{umari2009optimal}%
  \BibitemOpen
  \bibfield  {author} {\bibinfo {author} {\bibfnamefont {P.}~\bibnamefont
  {Umari}}, \bibinfo {author} {\bibfnamefont {G.}~\bibnamefont {Stenuit}}, \
  and\ \bibinfo {author} {\bibfnamefont {S.}~\bibnamefont {Baroni}},\
  }\href@noop {} {\bibfield  {journal} {\bibinfo  {journal} {Phys. Rev. B}\
  }\textbf {\bibinfo {volume} {79}},\ \bibinfo {pages} {201104} (\bibinfo
  {year} {2009})}\BibitemShut {NoStop}%
\bibitem [{\citenamefont {Umari}\ \emph {et~al.}(2010)\citenamefont {Umari},
  \citenamefont {Stenuit},\ and\ \citenamefont {Baroni}}]{umari2010gw}%
  \BibitemOpen
  \bibfield  {author} {\bibinfo {author} {\bibfnamefont {P.}~\bibnamefont
  {Umari}}, \bibinfo {author} {\bibfnamefont {G.}~\bibnamefont {Stenuit}}, \
  and\ \bibinfo {author} {\bibfnamefont {S.}~\bibnamefont {Baroni}},\
  }\href@noop {} {\bibfield  {journal} {\bibinfo  {journal} {Phys. Rev. B}\
  }\textbf {\bibinfo {volume} {81}},\ \bibinfo {pages} {115104} (\bibinfo
  {year} {2010})}\BibitemShut {NoStop}%
\bibitem [{\citenamefont {Deslippe}\ \emph {et~al.}(2012)\citenamefont
  {Deslippe}, \citenamefont {Samsonidze}, \citenamefont {Strubbe},
  \citenamefont {Jain}, \citenamefont {Cohen},\ and\ \citenamefont
  {Louie}}]{DeslippeSamsonidzeStrubbeEtAl2012}%
  \BibitemOpen
  \bibfield  {author} {\bibinfo {author} {\bibfnamefont {J.}~\bibnamefont
  {Deslippe}}, \bibinfo {author} {\bibfnamefont {G.}~\bibnamefont
  {Samsonidze}}, \bibinfo {author} {\bibfnamefont {D.~A.}\ \bibnamefont
  {Strubbe}}, \bibinfo {author} {\bibfnamefont {M.}~\bibnamefont {Jain}},
  \bibinfo {author} {\bibfnamefont {M.~L.}\ \bibnamefont {Cohen}}, \ and\
  \bibinfo {author} {\bibfnamefont {S.~G.}\ \bibnamefont {Louie}},\ }\href@noop
  {} {\bibfield  {journal} {\bibinfo  {journal} {Comput. Phys. Commun.}\
  }\textbf {\bibinfo {volume} {183}},\ \bibinfo {pages} {1269} (\bibinfo {year}
  {2012})}\BibitemShut {NoStop}%
\bibitem [{\citenamefont {Ljungberg}\ \emph {et~al.}(2015)\citenamefont
  {Ljungberg}, \citenamefont {Koval}, \citenamefont {Ferrari}, \citenamefont
  {Foerster},\ and\ \citenamefont {Sanchez-Portal}}]{ljungberg2015cubic}%
  \BibitemOpen
  \bibfield  {author} {\bibinfo {author} {\bibfnamefont {M.~P.}\ \bibnamefont
  {Ljungberg}}, \bibinfo {author} {\bibfnamefont {P.}~\bibnamefont {Koval}},
  \bibinfo {author} {\bibfnamefont {F.}~\bibnamefont {Ferrari}}, \bibinfo
  {author} {\bibfnamefont {D.}~\bibnamefont {Foerster}}, \ and\ \bibinfo
  {author} {\bibfnamefont {D.}~\bibnamefont {Sanchez-Portal}},\ }\href@noop {}
  {\bibfield  {journal} {\bibinfo  {journal} {Phys. Rev. B}\ }\textbf {\bibinfo
  {volume} {92}},\ \bibinfo {pages} {075422} (\bibinfo {year}
  {2015})}\BibitemShut {NoStop}%
\bibitem [{\citenamefont {Govoni}\ and\ \citenamefont
  {Galli}(2015)}]{govoni2015large}%
  \BibitemOpen
  \bibfield  {author} {\bibinfo {author} {\bibfnamefont {M.}~\bibnamefont
  {Govoni}}\ and\ \bibinfo {author} {\bibfnamefont {G.}~\bibnamefont {Galli}},\
  }\href@noop {} {\bibfield  {journal} {\bibinfo  {journal} {J. Chem. Theory
  Comput.}\ }\textbf {\bibinfo {volume} {11}},\ \bibinfo {pages} {2680}
  (\bibinfo {year} {2015})}\BibitemShut {NoStop}%
\bibitem [{\citenamefont {Wilhelm}\ \emph {et~al.}(2018)\citenamefont
  {Wilhelm}, \citenamefont {Golze}, \citenamefont {Talirz}, \citenamefont
  {Hutter},\ and\ \citenamefont {Pignedoli}}]{wilhelm2018toward}%
  \BibitemOpen
  \bibfield  {author} {\bibinfo {author} {\bibfnamefont {J.}~\bibnamefont
  {Wilhelm}}, \bibinfo {author} {\bibfnamefont {D.}~\bibnamefont {Golze}},
  \bibinfo {author} {\bibfnamefont {L.}~\bibnamefont {Talirz}}, \bibinfo
  {author} {\bibfnamefont {J.}~\bibnamefont {Hutter}}, \ and\ \bibinfo {author}
  {\bibfnamefont {C.~A.}\ \bibnamefont {Pignedoli}},\ }\href@noop {} {\bibfield
   {journal} {\bibinfo  {journal} {The journal of physical chemistry letters}\
  }\textbf {\bibinfo {volume} {9}},\ \bibinfo {pages} {306} (\bibinfo {year}
  {2018})}\BibitemShut {NoStop}%
\bibitem [{\citenamefont {Neuhauser}\ \emph
  {et~al.}(2014{\natexlab{a}})\citenamefont {Neuhauser}, \citenamefont {Gao},
  \citenamefont {Arntsen}, \citenamefont {Karshenas}, \citenamefont {Rabani},\
  and\ \citenamefont {Baer}}]{Neuhauser2014}%
  \BibitemOpen
  \bibfield  {author} {\bibinfo {author} {\bibfnamefont {D.}~\bibnamefont
  {Neuhauser}}, \bibinfo {author} {\bibfnamefont {Y.}~\bibnamefont {Gao}},
  \bibinfo {author} {\bibfnamefont {C.}~\bibnamefont {Arntsen}}, \bibinfo
  {author} {\bibfnamefont {C.}~\bibnamefont {Karshenas}}, \bibinfo {author}
  {\bibfnamefont {E.}~\bibnamefont {Rabani}}, \ and\ \bibinfo {author}
  {\bibfnamefont {R.}~\bibnamefont {Baer}},\ }\href {\doibase
  10.1103/PhysRevLett.113.076402} {\bibfield  {journal} {\bibinfo  {journal}
  {Phys. Rev. Lett.}\ }\textbf {\bibinfo {volume} {113}},\ \bibinfo {pages}
  {076402} (\bibinfo {year} {2014}{\natexlab{a}})}\BibitemShut {NoStop}%
\bibitem [{\citenamefont {Vl\v{c}ek}\ \emph {et~al.}(2017)\citenamefont
  {Vl\v{c}ek}, \citenamefont {Rabani}, \citenamefont {Neuhauser},\ and\
  \citenamefont {Baer}}]{vlcek2017stochastic}%
  \BibitemOpen
  \bibfield  {author} {\bibinfo {author} {\bibfnamefont {V.}~\bibnamefont
  {Vl\v{c}ek}}, \bibinfo {author} {\bibfnamefont {E.}~\bibnamefont {Rabani}},
  \bibinfo {author} {\bibfnamefont {D.}~\bibnamefont {Neuhauser}}, \ and\
  \bibinfo {author} {\bibfnamefont {R.}~\bibnamefont {Baer}},\ }\href@noop {}
  {\bibfield  {journal} {\bibinfo  {journal} {J. Chem. Theory Comput.}\
  }\textbf {\bibinfo {volume} {13}},\ \bibinfo {pages} {4997} (\bibinfo {year}
  {2017})}\BibitemShut {NoStop}%
\bibitem [{\citenamefont {Vl{\v{c}}ek}\ \emph {et~al.}(2016)\citenamefont
  {Vl{\v{c}}ek}, \citenamefont {Eisenberg}, \citenamefont {Steinle-Neumann},
  \citenamefont {Neuhauser}, \citenamefont {Rabani},\ and\ \citenamefont
  {Baer}}]{Vlcek2016}%
  \BibitemOpen
  \bibfield  {author} {\bibinfo {author} {\bibfnamefont {V.}~\bibnamefont
  {Vl{\v{c}}ek}}, \bibinfo {author} {\bibfnamefont {H.~R.}\ \bibnamefont
  {Eisenberg}}, \bibinfo {author} {\bibfnamefont {G.}~\bibnamefont
  {Steinle-Neumann}}, \bibinfo {author} {\bibfnamefont {D.}~\bibnamefont
  {Neuhauser}}, \bibinfo {author} {\bibfnamefont {E.}~\bibnamefont {Rabani}}, \
  and\ \bibinfo {author} {\bibfnamefont {R.}~\bibnamefont {Baer}},\ }\href@noop
  {} {\bibfield  {journal} {\bibinfo  {journal} {Phys. Rev. Lett.}\ }\textbf
  {\bibinfo {volume} {116}},\ \bibinfo {pages} {186401} (\bibinfo {year}
  {2016})}\BibitemShut {NoStop}%
\bibitem [{\citenamefont {Vl{\v{c}}ek}\ \emph {et~al.}(2018)\citenamefont
  {Vl{\v{c}}ek}, \citenamefont {Rabani},\ and\ \citenamefont
  {Neuhauser}}]{vlcek2018quasiparticle}%
  \BibitemOpen
  \bibfield  {author} {\bibinfo {author} {\bibfnamefont {V.}~\bibnamefont
  {Vl{\v{c}}ek}}, \bibinfo {author} {\bibfnamefont {E.}~\bibnamefont {Rabani}},
  \ and\ \bibinfo {author} {\bibfnamefont {D.}~\bibnamefont {Neuhauser}},\
  }\href@noop {} {\bibfield  {journal} {\bibinfo  {journal} {Phys Rev Mater}\
  }\textbf {\bibinfo {volume} {2}},\ \bibinfo {pages} {030801} (\bibinfo {year}
  {2018})}\BibitemShut {NoStop}%
\bibitem [{Note1()}]{Note1}%
  \BibitemOpen
  \bibinfo {note} {Note that generally quantities in time and frequency use the
  same symbol, and the specifics are clear from the argument.}\BibitemShut
  {Stop}%
\bibitem [{\citenamefont {Hedin}(1999)}]{hedin1999correlation}%
  \BibitemOpen
  \bibfield  {author} {\bibinfo {author} {\bibfnamefont {L.}~\bibnamefont
  {Hedin}},\ }\href@noop {} {\bibfield  {journal} {\bibinfo  {journal} {Journal
  of Physics: Condensed Matter}\ }\textbf {\bibinfo {volume} {11}},\ \bibinfo
  {pages} {R489} (\bibinfo {year} {1999})}\BibitemShut {NoStop}%
\bibitem [{\citenamefont {Runge}\ and\ \citenamefont
  {Gross}(1984)}]{runge1984density}%
  \BibitemOpen
  \bibfield  {author} {\bibinfo {author} {\bibfnamefont {E.}~\bibnamefont
  {Runge}}\ and\ \bibinfo {author} {\bibfnamefont {E.~K.}\ \bibnamefont
  {Gross}},\ }\href@noop {} {\bibfield  {journal} {\bibinfo  {journal} {Phys.
  Rev. Lett.}\ }\textbf {\bibinfo {volume} {52}},\ \bibinfo {pages} {997}
  (\bibinfo {year} {1984})}\BibitemShut {NoStop}%
\bibitem [{\citenamefont {Del~Sole}\ \emph {et~al.}(1994)\citenamefont
  {Del~Sole}, \citenamefont {Reining},\ and\ \citenamefont
  {Godby}}]{del1994gwgamma}%
  \BibitemOpen
  \bibfield  {author} {\bibinfo {author} {\bibfnamefont {R.}~\bibnamefont
  {Del~Sole}}, \bibinfo {author} {\bibfnamefont {L.}~\bibnamefont {Reining}}, \
  and\ \bibinfo {author} {\bibfnamefont {R.}~\bibnamefont {Godby}},\
  }\href@noop {} {\bibfield  {journal} {\bibinfo  {journal} {Phys. Rev. B}\
  }\textbf {\bibinfo {volume} {49}},\ \bibinfo {pages} {8024} (\bibinfo {year}
  {1994})}\BibitemShut {NoStop}%
\bibitem [{\citenamefont {Bruneval}\ \emph {et~al.}(2005)\citenamefont
  {Bruneval}, \citenamefont {Sottile}, \citenamefont {Olevano}, \citenamefont
  {Del~Sole},\ and\ \citenamefont {Reining}}]{bruneval2005many}%
  \BibitemOpen
  \bibfield  {author} {\bibinfo {author} {\bibfnamefont {F.}~\bibnamefont
  {Bruneval}}, \bibinfo {author} {\bibfnamefont {F.}~\bibnamefont {Sottile}},
  \bibinfo {author} {\bibfnamefont {V.}~\bibnamefont {Olevano}}, \bibinfo
  {author} {\bibfnamefont {R.}~\bibnamefont {Del~Sole}}, \ and\ \bibinfo
  {author} {\bibfnamefont {L.}~\bibnamefont {Reining}},\ }\href@noop {}
  {\bibfield  {journal} {\bibinfo  {journal} {Phys. Rev. Lett.}\ }\textbf
  {\bibinfo {volume} {94}},\ \bibinfo {pages} {186402} (\bibinfo {year}
  {2005})}\BibitemShut {NoStop}%
\bibitem [{\citenamefont {Gr{\"u}neis}\ \emph {et~al.}(2014)\citenamefont
  {Gr{\"u}neis}, \citenamefont {Kresse}, \citenamefont {Hinuma},\ and\
  \citenamefont {Oba}}]{gruneis2014ionization}%
  \BibitemOpen
  \bibfield  {author} {\bibinfo {author} {\bibfnamefont {A.}~\bibnamefont
  {Gr{\"u}neis}}, \bibinfo {author} {\bibfnamefont {G.}~\bibnamefont {Kresse}},
  \bibinfo {author} {\bibfnamefont {Y.}~\bibnamefont {Hinuma}}, \ and\ \bibinfo
  {author} {\bibfnamefont {F.}~\bibnamefont {Oba}},\ }\href@noop {} {\bibfield
  {journal} {\bibinfo  {journal} {Phys. Rev. Lett.}\ }\textbf {\bibinfo
  {volume} {112}},\ \bibinfo {pages} {096401} (\bibinfo {year}
  {2014})}\BibitemShut {NoStop}%
\bibitem [{\citenamefont {Hung}\ \emph {et~al.}(2016)\citenamefont {Hung},
  \citenamefont {da~Jornada}, \citenamefont {Souto-Casares}, \citenamefont
  {Chelikowsky}, \citenamefont {Louie},\ and\ \citenamefont
  {{\"O}{\u{g}}{\"u}t}}]{hung2016excitation}%
  \BibitemOpen
  \bibfield  {author} {\bibinfo {author} {\bibfnamefont {L.}~\bibnamefont
  {Hung}}, \bibinfo {author} {\bibfnamefont {F.~H.}\ \bibnamefont
  {da~Jornada}}, \bibinfo {author} {\bibfnamefont {J.}~\bibnamefont
  {Souto-Casares}}, \bibinfo {author} {\bibfnamefont {J.~R.}\ \bibnamefont
  {Chelikowsky}}, \bibinfo {author} {\bibfnamefont {S.~G.}\ \bibnamefont
  {Louie}}, \ and\ \bibinfo {author} {\bibfnamefont {S.}~\bibnamefont
  {{\"O}{\u{g}}{\"u}t}},\ }\href@noop {} {\bibfield  {journal} {\bibinfo
  {journal} {Phys. Rev. B}\ }\textbf {\bibinfo {volume} {94}},\ \bibinfo
  {pages} {085125} (\bibinfo {year} {2016})}\BibitemShut {NoStop}%
\bibitem [{\citenamefont {Vl{\v{c}}ek}\ \emph {et~al.}(2017)\citenamefont
  {Vl{\v{c}}ek}, \citenamefont {Baer}, \citenamefont {Rabani},\ and\
  \citenamefont {Neuhauser}}]{vlcek2017simple}%
  \BibitemOpen
  \bibfield  {author} {\bibinfo {author} {\bibfnamefont {V.}~\bibnamefont
  {Vl{\v{c}}ek}}, \bibinfo {author} {\bibfnamefont {R.}~\bibnamefont {Baer}},
  \bibinfo {author} {\bibfnamefont {E.}~\bibnamefont {Rabani}}, \ and\ \bibinfo
  {author} {\bibfnamefont {D.}~\bibnamefont {Neuhauser}},\ }\href@noop {}
  {\bibfield  {journal} {\bibinfo  {journal} {arXiv preprint arXiv:1701.02023}\
  } (\bibinfo {year} {2017})}\BibitemShut {NoStop}%
\bibitem [{\citenamefont {Gao}\ \emph {et~al.}(2015)\citenamefont {Gao},
  \citenamefont {Neuhauser}, \citenamefont {Baer},\ and\ \citenamefont
  {Rabani}}]{Rabani2015}%
  \BibitemOpen
  \bibfield  {author} {\bibinfo {author} {\bibfnamefont {Y.}~\bibnamefont
  {Gao}}, \bibinfo {author} {\bibfnamefont {D.}~\bibnamefont {Neuhauser}},
  \bibinfo {author} {\bibfnamefont {R.}~\bibnamefont {Baer}}, \ and\ \bibinfo
  {author} {\bibfnamefont {E.}~\bibnamefont {Rabani}},\ }\href@noop {}
  {\bibfield  {journal} {\bibinfo  {journal} {J. Chem. Phys.}\ }\textbf
  {\bibinfo {volume} {142}},\ \bibinfo {pages} {034106} (\bibinfo {year}
  {2015})}\BibitemShut {NoStop}%
\bibitem [{\citenamefont {Rabani}\ \emph {et~al.}(2015)\citenamefont {Rabani},
  \citenamefont {Neuhauser},\ and\ \citenamefont {Baer}}]{Rabani2015a}%
  \BibitemOpen
  \bibfield  {author} {\bibinfo {author} {\bibfnamefont {E.}~\bibnamefont
  {Rabani}}, \bibinfo {author} {\bibfnamefont {D.}~\bibnamefont {Neuhauser}}, \
  and\ \bibinfo {author} {\bibfnamefont {R.}~\bibnamefont {Baer}},\ }\href@noop
  {} {\bibfield  {journal} {\bibinfo  {journal} {Phys. Rev. B}\ }\textbf
  {\bibinfo {volume} {91}},\ \bibinfo {pages} {235302} (\bibinfo {year}
  {2015})}\BibitemShut {NoStop}%
\bibitem [{\citenamefont {Neuhauser}\ \emph {et~al.}(2015)\citenamefont
  {Neuhauser}, \citenamefont {Rabani}, \citenamefont {Cytter},\ and\
  \citenamefont {Baer}}]{neuhauser2015stochastic}%
  \BibitemOpen
  \bibfield  {author} {\bibinfo {author} {\bibfnamefont {D.}~\bibnamefont
  {Neuhauser}}, \bibinfo {author} {\bibfnamefont {E.}~\bibnamefont {Rabani}},
  \bibinfo {author} {\bibfnamefont {Y.}~\bibnamefont {Cytter}}, \ and\ \bibinfo
  {author} {\bibfnamefont {R.}~\bibnamefont {Baer}},\ }\href@noop {} {\bibfield
   {journal} {\bibinfo  {journal} {J. Phys. Chem. A}\ }\textbf {\bibinfo
  {volume} {120}},\ \bibinfo {pages} {3071} (\bibinfo {year}
  {2015})}\BibitemShut {NoStop}%
\bibitem [{\citenamefont {Fetter}\ and\ \citenamefont
  {Walecka}(1971)}]{Fetter1971}%
  \BibitemOpen
  \bibfield  {author} {\bibinfo {author} {\bibfnamefont {A.~L.}\ \bibnamefont
  {Fetter}}\ and\ \bibinfo {author} {\bibfnamefont {J.~D.}\ \bibnamefont
  {Walecka}},\ }\href@noop {} {\emph {\bibinfo {title} {Quantum Thoery of Many
  Particle Systems}}}\ (\bibinfo  {publisher} {McGraw-Hill},\ \bibinfo
  {address} {New York},\ \bibinfo {year} {1971})\ p.\ \bibinfo {pages}
  {299}\BibitemShut {NoStop}%
\bibitem [{Note2()}]{Note2}%
  \BibitemOpen
  \bibinfo {note} {If a segment starting point is chosen near the first or last
  point in the grid, then either the function should be wrapped (so a portion
  of the segment is near the end of the grid and another portion is near the
  beginning of the grid) or it should be padded (at the beginning or end) with
  zeros, to guarantee that all points are equally sampled.}\BibitemShut {Stop}%
\bibitem [{\citenamefont {Troullier}\ and\ \citenamefont
  {Martins}(1991)}]{TroullierMartins1991}%
  \BibitemOpen
  \bibfield  {author} {\bibinfo {author} {\bibfnamefont {N.}~\bibnamefont
  {Troullier}}\ and\ \bibinfo {author} {\bibfnamefont {J.~L.}\ \bibnamefont
  {Martins}},\ }\href@noop {} {\bibfield  {journal} {\bibinfo  {journal} {Phys.
  Rev. B}\ }\textbf {\bibinfo {volume} {43}},\ \bibinfo {pages} {1993}
  (\bibinfo {year} {1991})}\BibitemShut {NoStop}%
\bibitem [{\citenamefont {Martyna}\ and\ \citenamefont
  {Tuckerman}(1999)}]{martyna1999reciprocal}%
  \BibitemOpen
  \bibfield  {author} {\bibinfo {author} {\bibfnamefont {G.~J.}\ \bibnamefont
  {Martyna}}\ and\ \bibinfo {author} {\bibfnamefont {M.~E.}\ \bibnamefont
  {Tuckerman}},\ }\href@noop {} {\bibfield  {journal} {\bibinfo  {journal} {J.
  Chem. Phys.}\ }\textbf {\bibinfo {volume} {110}},\ \bibinfo {pages} {2810}
  (\bibinfo {year} {1999})}\BibitemShut {NoStop}%
\bibitem [{\citenamefont {Yamanaka}\ and\ \citenamefont
  {Morimoto}(1996)}]{yamanaka1996isotope}%
  \BibitemOpen
  \bibfield  {author} {\bibinfo {author} {\bibfnamefont {T.}~\bibnamefont
  {Yamanaka}}\ and\ \bibinfo {author} {\bibfnamefont {S.}~\bibnamefont
  {Morimoto}},\ }\href@noop {} {\bibfield  {journal} {\bibinfo  {journal} {Acta
  Crystallogr., Sect. B: Struct. Sci}\ }\textbf {\bibinfo {volume} {52}},\
  \bibinfo {pages} {232} (\bibinfo {year} {1996})}\BibitemShut {NoStop}%
\bibitem [{\citenamefont {Elliot}(2010)}]{elliot2010structure}%
  \BibitemOpen
  \bibfield  {author} {\bibinfo {author} {\bibfnamefont {A.~D.}\ \bibnamefont
  {Elliot}},\ }\href@noop {} {\bibfield  {journal} {\bibinfo  {journal} {Acta
  Crystallogr., Sect. B: Struct. Sci}\ }\textbf {\bibinfo {volume} {66}},\
  \bibinfo {pages} {271} (\bibinfo {year} {2010})}\BibitemShut {NoStop}%
\bibitem [{Note3()}]{Note3}%
  \BibitemOpen
  \bibinfo {note} {The StochasticGW code is available under GPL at
  http://www.stochasticgw.com}\BibitemShut {NoStop}%
\bibitem [{\citenamefont {Yang}(1991)}]{yang1991direct}%
  \BibitemOpen
  \bibfield  {author} {\bibinfo {author} {\bibfnamefont {W.}~\bibnamefont
  {Yang}},\ }\href@noop {} {\bibfield  {journal} {\bibinfo  {journal} {Physical
  Review Letters}\ }\textbf {\bibinfo {volume} {66}},\ \bibinfo {pages} {1438}
  (\bibinfo {year} {1991})}\BibitemShut {NoStop}%
\bibitem [{\citenamefont {Hern{\'a}ndez}\ and\ \citenamefont
  {Gillan}(1995)}]{hernandez1995self}%
  \BibitemOpen
  \bibfield  {author} {\bibinfo {author} {\bibfnamefont {E.}~\bibnamefont
  {Hern{\'a}ndez}}\ and\ \bibinfo {author} {\bibfnamefont {M.}~\bibnamefont
  {Gillan}},\ }\href@noop {} {\bibfield  {journal} {\bibinfo  {journal}
  {Physical Review B}\ }\textbf {\bibinfo {volume} {51}},\ \bibinfo {pages}
  {10157} (\bibinfo {year} {1995})}\BibitemShut {NoStop}%
\bibitem [{\citenamefont {Mohr}\ \emph {et~al.}(2015)\citenamefont {Mohr},
  \citenamefont {Ratcliff}, \citenamefont {Genovese}, \citenamefont {Caliste},
  \citenamefont {Boulanger}, \citenamefont {Goedecker},\ and\ \citenamefont
  {Deutsch}}]{mohr2015accurate}%
  \BibitemOpen
  \bibfield  {author} {\bibinfo {author} {\bibfnamefont {S.}~\bibnamefont
  {Mohr}}, \bibinfo {author} {\bibfnamefont {L.~E.}\ \bibnamefont {Ratcliff}},
  \bibinfo {author} {\bibfnamefont {L.}~\bibnamefont {Genovese}}, \bibinfo
  {author} {\bibfnamefont {D.}~\bibnamefont {Caliste}}, \bibinfo {author}
  {\bibfnamefont {P.}~\bibnamefont {Boulanger}}, \bibinfo {author}
  {\bibfnamefont {S.}~\bibnamefont {Goedecker}}, \ and\ \bibinfo {author}
  {\bibfnamefont {T.}~\bibnamefont {Deutsch}},\ }\href@noop {} {\bibfield
  {journal} {\bibinfo  {journal} {Physical Chemistry Chemical Physics}\
  }\textbf {\bibinfo {volume} {17}},\ \bibinfo {pages} {31360} (\bibinfo {year}
  {2015})}\BibitemShut {NoStop}%
\bibitem [{\citenamefont {VandeVondele}\ \emph {et~al.}(2012)\citenamefont
  {VandeVondele}, \citenamefont {Borstnik},\ and\ \citenamefont
  {Hutter}}]{vandevondele2012linear}%
  \BibitemOpen
  \bibfield  {author} {\bibinfo {author} {\bibfnamefont {J.}~\bibnamefont
  {VandeVondele}}, \bibinfo {author} {\bibfnamefont {U.}~\bibnamefont
  {Borstnik}}, \ and\ \bibinfo {author} {\bibfnamefont {J.}~\bibnamefont
  {Hutter}},\ }\href@noop {} {\bibfield  {journal} {\bibinfo  {journal}
  {Journal of chemical theory and computation}\ }\textbf {\bibinfo {volume}
  {8}},\ \bibinfo {pages} {3565} (\bibinfo {year} {2012})}\BibitemShut
  {NoStop}%
\bibitem [{\citenamefont {Baer}\ \emph {et~al.}(2013)\citenamefont {Baer},
  \citenamefont {Neuhauser},\ and\ \citenamefont
  {Rabani}}]{BaerNeuhauserRabani2013}%
  \BibitemOpen
  \bibfield  {author} {\bibinfo {author} {\bibfnamefont {R.}~\bibnamefont
  {Baer}}, \bibinfo {author} {\bibfnamefont {D.}~\bibnamefont {Neuhauser}}, \
  and\ \bibinfo {author} {\bibfnamefont {E.}~\bibnamefont {Rabani}},\
  }\href@noop {} {\bibfield  {journal} {\bibinfo  {journal} {Phys. Rev. Lett.}\
  }\textbf {\bibinfo {volume} {111}},\ \bibinfo {pages} {106402} (\bibinfo
  {year} {2013})}\BibitemShut {NoStop}%
\bibitem [{\citenamefont {Neuhauser}\ \emph
  {et~al.}(2014{\natexlab{b}})\citenamefont {Neuhauser}, \citenamefont {Baer},\
  and\ \citenamefont {Rabani}}]{neuhauser2014communication}%
  \BibitemOpen
  \bibfield  {author} {\bibinfo {author} {\bibfnamefont {D.}~\bibnamefont
  {Neuhauser}}, \bibinfo {author} {\bibfnamefont {R.}~\bibnamefont {Baer}}, \
  and\ \bibinfo {author} {\bibfnamefont {E.}~\bibnamefont {Rabani}},\
  }\href@noop {} {\bibfield  {journal} {\bibinfo  {journal} {J. Chem. Phys.}\
  }\textbf {\bibinfo {volume} {141}},\ \bibinfo {pages} {041102} (\bibinfo
  {year} {2014}{\natexlab{b}})}\BibitemShut {NoStop}%
\bibitem [{\citenamefont {Neuhauser}\ \emph {et~al.}(2017)\citenamefont
  {Neuhauser}, \citenamefont {Baer},\ and\ \citenamefont
  {Zgid}}]{neuhauser2017stochastic}%
  \BibitemOpen
  \bibfield  {author} {\bibinfo {author} {\bibfnamefont {D.}~\bibnamefont
  {Neuhauser}}, \bibinfo {author} {\bibfnamefont {R.}~\bibnamefont {Baer}}, \
  and\ \bibinfo {author} {\bibfnamefont {D.}~\bibnamefont {Zgid}},\ }\href@noop
  {} {\bibfield  {journal} {\bibinfo  {journal} {J. Chem. Theory Comput.}\
  }\textbf {\bibinfo {volume} {13}},\ \bibinfo {pages} {5396} (\bibinfo {year}
  {2017})}\BibitemShut {NoStop}%
\bibitem [{\citenamefont {Takeshita}\ \emph {et~al.}(2017)\citenamefont
  {Takeshita}, \citenamefont {de~Jong}, \citenamefont {Neuhauser},
  \citenamefont {Baer},\ and\ \citenamefont
  {Rabani}}]{takeshita2017stochastic}%
  \BibitemOpen
  \bibfield  {author} {\bibinfo {author} {\bibfnamefont {T.~Y.}\ \bibnamefont
  {Takeshita}}, \bibinfo {author} {\bibfnamefont {W.~A.}\ \bibnamefont
  {de~Jong}}, \bibinfo {author} {\bibfnamefont {D.}~\bibnamefont {Neuhauser}},
  \bibinfo {author} {\bibfnamefont {R.}~\bibnamefont {Baer}}, \ and\ \bibinfo
  {author} {\bibfnamefont {E.}~\bibnamefont {Rabani}},\ }\href@noop {}
  {\bibfield  {journal} {\bibinfo  {journal} {J. Chem. Theory Comput.}\
  }\textbf {\bibinfo {volume} {13}},\ \bibinfo {pages} {4605} (\bibinfo {year}
  {2017})}\BibitemShut {NoStop}%
\bibitem [{\citenamefont {Towns}\ \emph {et~al.}(2014)\citenamefont {Towns},
  \citenamefont {Cockerill}, \citenamefont {Dahan}, \citenamefont {Foster},
  \citenamefont {Gaither}, \citenamefont {Grimshaw}, \citenamefont {Hazlewood},
  \citenamefont {Lathrop}, \citenamefont {Lifka}, \citenamefont {Peterson}
  \emph {et~al.}}]{towns2014xsede}%
  \BibitemOpen
  \bibfield  {author} {\bibinfo {author} {\bibfnamefont {J.}~\bibnamefont
  {Towns}}, \bibinfo {author} {\bibfnamefont {T.}~\bibnamefont {Cockerill}},
  \bibinfo {author} {\bibfnamefont {M.}~\bibnamefont {Dahan}}, \bibinfo
  {author} {\bibfnamefont {I.}~\bibnamefont {Foster}}, \bibinfo {author}
  {\bibfnamefont {K.}~\bibnamefont {Gaither}}, \bibinfo {author} {\bibfnamefont
  {A.}~\bibnamefont {Grimshaw}}, \bibinfo {author} {\bibfnamefont
  {V.}~\bibnamefont {Hazlewood}}, \bibinfo {author} {\bibfnamefont
  {S.}~\bibnamefont {Lathrop}}, \bibinfo {author} {\bibfnamefont
  {D.}~\bibnamefont {Lifka}}, \bibinfo {author} {\bibfnamefont {G.~D.}\
  \bibnamefont {Peterson}},  \emph {et~al.},\ }\href@noop {} {\bibfield
  {journal} {\bibinfo  {journal} {Computing in Science \& Engineering}\
  }\textbf {\bibinfo {volume} {16}},\ \bibinfo {pages} {62} (\bibinfo {year}
  {2014})}\BibitemShut {NoStop}%
\end{thebibliography}%

\end{document}